\documentclass{PoS}

\usepackage[T1]{fontenc}
\usepackage{graphicx}
\usepackage{subfigure}
\usepackage{amsmath}
\usepackage{amssymb}

\title{Hard Exclusive $\Phi$ and $J/\Psi$ Photoproduction off a Proton}

\ShortTitle{Hard Exclusive $\Phi$ and $J/\Psi$ Photoproduction off a Proton}

\author{\speaker{A.\ T.\ Goritschnig} \\ 
        Institute of Physics - Theory Division, University of Graz, Graz, Austria \\
        E-mail: \email{alexander.goritschnig@uni-graz.at}}

\author{B.\ Meli{\'c} \\
        Ru{\dj}er Bo\v skovi{\'c} Institute - Division of Theoretical Physics, Zagreb, Croatia \\
        E-mail: \email{melic@irb.hr}}
        
\author{K.\ Passek-Kumeri\v cki \\
        Ru{\dj}er Bo\v skovi{\'c} Institute - Division of Theoretical Physics, Zagreb, Croatia \\
        E-mail: \email{passek@irb.hr}}
        
\author{W.\ Schweiger \\
        Institute of Physics - Theory Division, University of Graz, Graz, Austria \\
        E-mail: \email{wolfgang.schweiger@uni-graz.at}}

\abstract{
We study the photoproduction of the vector mesons $\Phi$ and $J/\Psi$ off a proton in the kinematical regime of large energies and scattering angles within the framework of perturbative QCD. Our investigations are based on the hard scattering approach. This means that the hadrons are replaced by their valence Fock states and scattering on the partonic level is described by tree graphs in which the large transferred momentum is redistributed between the valence partons via the exchange of hard gluons. We find that the unpolarized photoproduction cross sections are dominated by Compton-scattering-like graphs in which the photon couples to the proton, whereas vector-meson-dominance-like graphs, in which the photon fluctuates into the heavy quark-antiquark pair which then exchanges two gluons with the proton, play a minor role. We give explicit predictions for unpolarized scattering cross sections and compare them with experimental data where possible.
}

\FullConference{XXII. International Workshop on Deep-Inelastic Scattering and Related Subjects \\
                 28 April - 2 May 2014 \\
                 Warsaw, Poland}

\begin{document}

\section{Introduction}
We are interested in the perturabtive treatment of vector-meson photoproduction  
\begin{equation}
 \gamma (p_\gamma,\,\lambda_\gamma) \,\,  p (p_i,\,\lambda_i)
 \,\,\, \longrightarrow \,\,\, 
 M_V (p_{M_V},\,\lambda_{M_V}) \,\, p (p_f,\,\lambda_f)  
  \quad \text{where} \quad 
 M_V \,=\, \Phi \,\, \text{or} \,\, J/\Psi \,. 
\label{eq:ourprocess}
\end{equation}
As a prerequisite for the application of perturbative QCD (pQCD) the kinematical situation has to be 
such that the process under consideration is dominated by small space-time-distance effects. 
This requires the existence of a large energy scale, i.e., hard scale, in the process.
 
For our reaction there are, in principle, three energy scales at hand which can be large. 
These are the virtuality of the incoming photon, the heavy masses of the (anti)quarks inside the vector mesons 
and the (transverse) momentum transfer.  
Different orderings of magnitudes of these scales correspond to different kinematical situations. 
In our case the incoming photon virtuality is equal to zero since we consider photoproduction, i.e., a real incoming photon. 
The mass of the heavy (anti)quark inside the vector meson (especially the $c$-quark mass when a $J/\Psi$ is produced) 
provides also a large energy scale. 
But we want to concentrate rather on the situation where the four-momentum transfer becomes largest (of the order of a few $\text{GeV}$),
i.e., larger than the heavy (anti)quark mass.  

Having a hard scale at hand one then has to separate the hard short-distance from the soft long-distance dynamics; 
this goes under the name of factorization. 
The hard part can be dealt within perturbation theory by evaluating Feynman diagrams, 
whereas the soft one has to be treated by introducing a priori unknown non-perturbative functions. 
There are several factorization schemes which differ in the way this splitting is done, i.e.,  
what is attributed to the hard and the soft parts, respectively. 

We investigate our process within the factorization scheme of the, so-called, \lq\lq hard scattering approach\rq\rq\ (HSA) introduced in Refs.~\cite{ER, BL}. 
It is generally accepted that the HSA is the dominant reaction mechanism at asymptotically large momentum transfer. 
But it strongly depends on the particular reaction whether it also provides a substantial contribution at 
moderately large momentum transfer of the order of a few GeV.  
This chance only exists if competing mechanisms play a minor role for one or the other reason. 
This could be the case for the $\Phi$ and $J/\Psi$ photoproduction channels, 
since vector-meson-dominance as well as handbag-type mechanisms
are suppressed if a heavy quark-antiquark pair has to be produced.

\section{$\Phi$ and $J/\Psi$ Photoproduction within the Hard Scattering Approach}
\label{sec:HSA}
Within the HSA the hard part of the hadronic scattering amplitude is obtained by replacing each hadron by its valence Fock-state. 
The specific feature of the HSA is that there are no spectators. 
This is realized by introducing the minimal number of gluons to connect all  hadronic constituents that participate in the scattering process on the partonic level. 
These gluons take care of the internal redistribution of the large transferred momentum such that the probability that the hadrons stay intact is maximized.  Thus, within the HSA, the hard scattering is represented by a coherent sum of Feynman tree-diagrams.  
Furthermore, internal transverse momenta of the hadronic constituents are neglected in the hard part and the constituents are assumed to be nearly on-mass-shell. In this approximation the constituents move collinear to their parent hadron and their momenta are uniquely specified by the fractions of the parent hadron momenta they carry. 
In the HSA masses of the light (current) quarks are usually set to zero.
When we consider $J/\Psi$ production, however,  we take the (anti)charm-quark mass into account 
by setting it equal to half the $J/\Psi$ mass.   

\begin{figure}[htb]
\subfigure{\includegraphics[width=0.45\textwidth, height=0.35\textwidth]{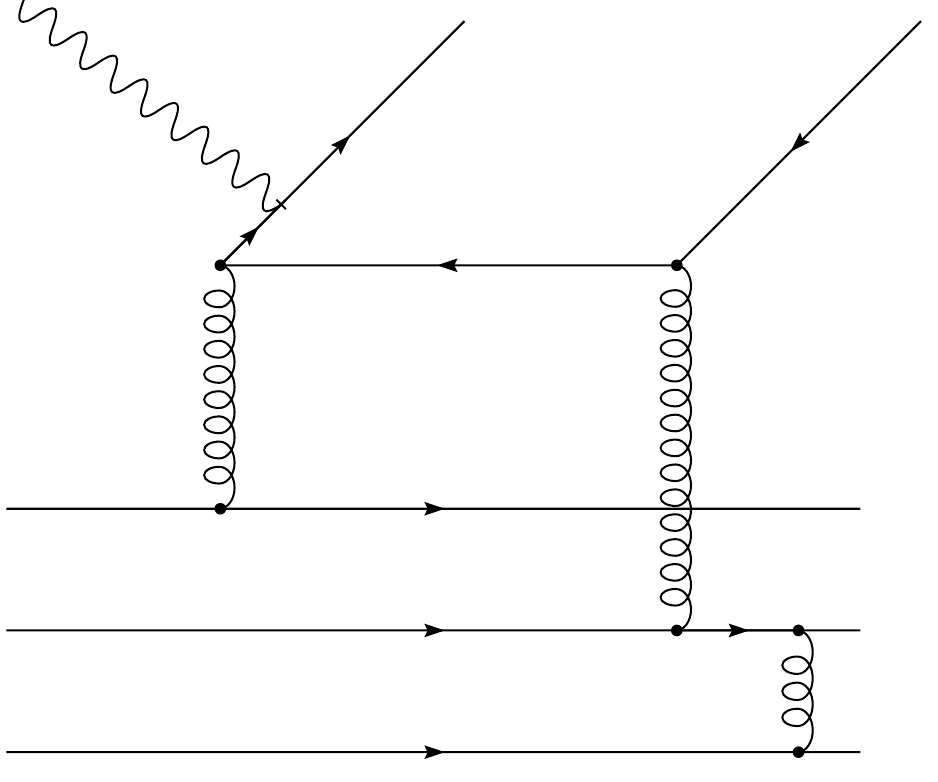}}
\hfill
\subfigure{\includegraphics[width=0.45\textwidth, height=0.35\textwidth]{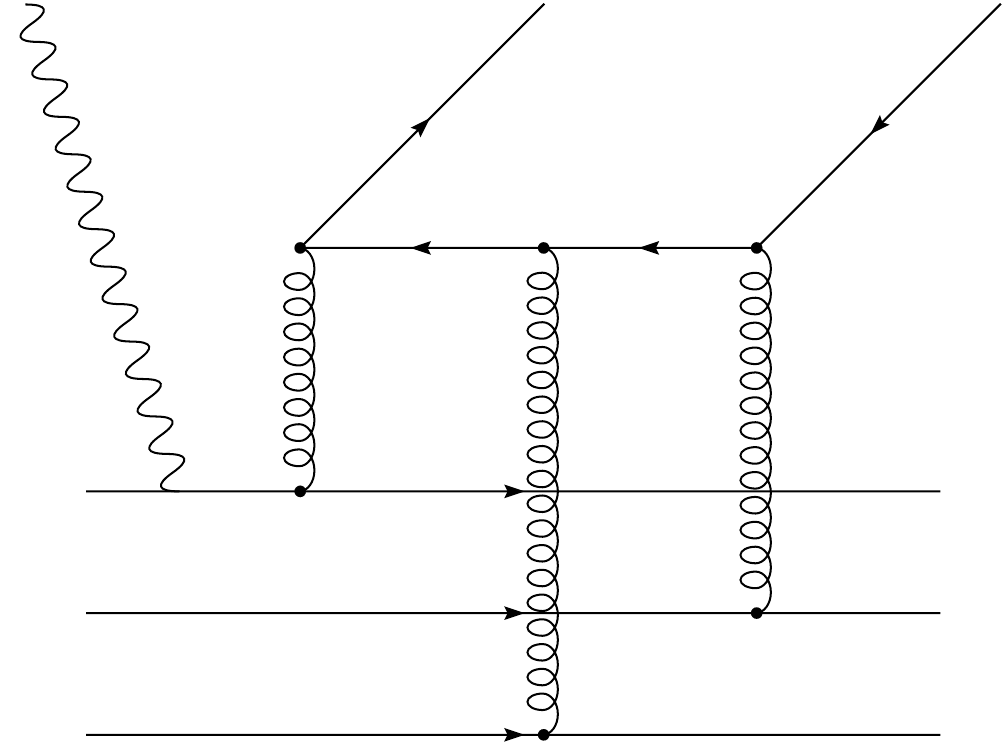}}
\caption{
 Two typical representatives of graphs, which contribute to the hard scattering amplitude for 
 vector meson photoproduction. 
 Graphs, such as the left one, where the photon (wavy line) couples to the vector meson, are  
 denoted as Class~I graphs. 
 Graphs having the structure of the right one are called Class~II. 
 Here the photon couples directly to the proton. 
Solid lines represent quarks and curled ones gluons.
} 
\label{classIandIV}
\end{figure}

The Feynman diagrams contributing to the (hard) scattering amplitude on the partonic level can be grouped into two generic classes as shown in Fig.~\ref{classIandIV}: 
one in which the photon couples to the vector meson (which we call Class~I) and 
one in which it couples to the proton (called Class~II). 
Class~I diagrams can be considered as a remnant of a vector-meson-dominance mechanism,
in which the photon fluctuates into the vector meson which then goes on-shell by exchanging hard gluons with the proton. 
Class~II diagrams, where the photon couples directly to the proton, represent a Compton-scattering-like mechanism.  
The soft, non-perturbative ingredients of the HSA are the hadron distribution amplitudes (DAs). 
These are probability amplitudes for finding the hadronic constituents in a certain momentum fraction configuration inside their parent hadron.  
As such they comprehend the bound-state dynamics of the valence partons of the incoming and outgoing hadrons. 
In our actual calculation we have chosen the ``asymptotic'' DA for the proton, $\phi_p \propto x_1 x_2 x_3$, and 
the \lq\lq non-relativistic\rq\rq\ DA $\phi_V \propto \delta\left(z_1- 1/2\right)$ for the vector mesons. 
According to the factorization scheme of the HSA the $\gamma\,p \,\to\, M_V\,p$ scattering amplitude $M$ is  
a convolution integral of a hard scattering amplitude $\hat{T}$ (i.e. the coherent sum of tree graphs) 
and the hadronic DAs $\phi_p(x_1,\, x_2,\, x_3)$, $\phi_p(y_1,\, y_2,\, y_3)$ and $\phi_{M_V}(z_1,\, z_2)$ for 
the incoming proton, the outgoing proton and the produced vector meson, respectively: 
\begin{equation} 
 M(s,t) = 
 \int_0^1\left[dx\right] \int_0^1\left[ dy\right] \int_0^1\left[ dz\right] \,
 \phi^{\dagger}_{M_V}\left(z_1, z_2\right)
 \phi^{\dagger}_p\left(y_1, y_2, y_3\right)
 \hat{T}\left(x_1,.., y_1,.., z_1.. ; s, t\right) 
 \phi_p\left(x_1, x_2, x_3\right)\,. 
\label{eq:convolution}
\end{equation}
The integration has to be done with respect to the partonic momentum fractions. 
The $x_i$ and $y_i$, $i=1,..,3$, denote the momentum fractions of the valence quarks inside the incoming and outgoing proton, respectively, 
and the $z_i$, $i=1,2$, those of the valence quark and antiquark inside the produced vector meson. 
The integration measures in Eq.~(\ref{eq:convolution}) are chosen such that the conditions 
$x_1+x_2+x_3 = 1$, $y_1+y_2+y_3 = 1$ and $z_1+z_2 = 1$ are fulfilled. 
The $\gamma\,p \,\to\, M_V\,p$ amplitude $M$ can then be expressed as a function of the two kinematical independent variables Mandelstam $s$ and $t$. 
The hard scattering amplitude $\hat{T}$ is a function of the Mandelstam variables $\hat{s}$ and $\hat{t}$ on partonic level, 
which can be related to $s$ and $t$ with the help of the partonic momentum fractions.

The numerical evaluation of the convolution integral in Eq.~(\ref{eq:convolution}) has to be done with care. 
The diagrams contributing to $\hat{T}$ exhibit propagator singularities which have to be treated properly. 
We have adopted the following procedure: 
Contributions coming from propagators which can become singular are first isolated by means of a partial fractioning. This has to be done for each of the, altogether, 108 diagrams.  The propagator denominators in the singular terms are then treated by means of the usual $\imath\epsilon$ prescription, 
\begin{equation}
 \frac{1}{k^2 \pm \imath\epsilon} = \mathcal{P}\left(\frac{1}{k^2}\right) \mp \imath\pi\delta\left(k^2\right) \, .
\label{eq:iepsilon}
\end{equation}
The integration of the delta-function part is trivial. The principal-value part is further split into an analytically solvable principal-value integral and a regular contribution, which can be treated with standard numerical integration routines.  

For the strong coupling $\alpha_{\mathrm S}$ we have used the one-loop expression.  
As argument we have taken the average virtuality of the hard gluon that gives rise to a particular $\alpha_{\mathrm S}$. 
If $|t|$, however, becomes too small, $\alpha_{\mathrm S}$ can become too large to justify a perturbative treatment. 
In order to extrapolate our predictions to small values of $|t|$ (where data exist), 
we apply a cutoff such that $\alpha_{\mathrm S}$ does not exceed a value of 0.7 for $\Phi$ and 0.5 for $J/\Psi$ production.

\section{Predictions for $\Phi$ and $J/\Psi$ Photoproduction}
\label{sec:results}
\begin{figure}[b]
\subfigure{\includegraphics[width=0.47\textwidth, height=0.35\textwidth]{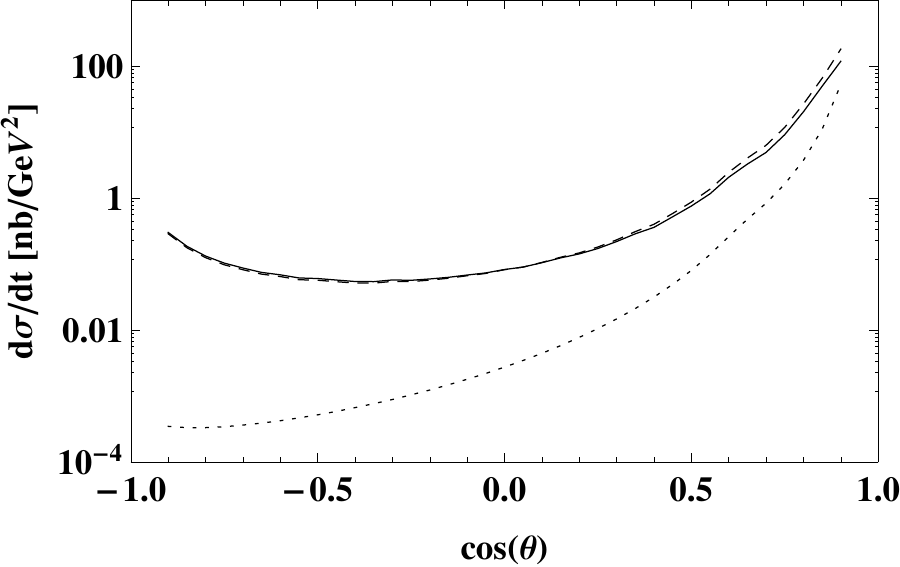}}
\hfill
\subfigure{\includegraphics[width=0.47\textwidth, height=0.35\textwidth]{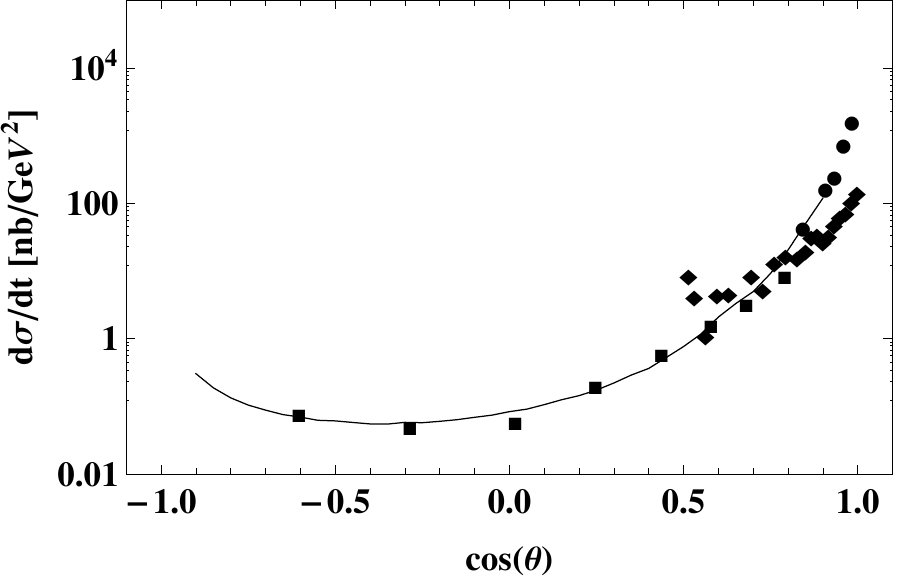}}
\caption{
Our prediction for $d\sigma_{\gamma p \rightarrow \Phi p}/dt$ at $p_\gamma^{(lab)}= 9.3$~GeV. 
The {\it left} panel exhibits the contributions of Class~I (dotted) and Class~II (dashed) diagrams to the full result (solid). 
On the {\it right} panel we compare our results with experimental data~\cite{Ballam,Barber,Anciant} ($\bullet$, $\blacktriangle$, $\blacksquare$).
Data from Refs.~\cite{Barber,Anciant} are appropriately scaled up to $p_\gamma^\mathrm{lab}= 9.3$~GeV. 
} 
\label{fig:Phiproduction}
\end{figure}

In Fig.~\ref{fig:Phiproduction} we present our predictions for the unpolarized differential 
$\gamma\,p \,\to\, \Phi\,p$ cross section at the photon lab energy $p_\gamma^{(lab)} = 9.3\,\text{GeV}$ 
(corresponding to $s = 18.33\,\text{GeV}^2$). 
The left panel of Fig.~\ref{fig:Phiproduction} shows how Class~I (dotted line) and Class~II (dashed line) graphs contribute to  
the full (solid line) unpolarized differential cross section. 
One observes that the Compton-scattering-like Class~II diagrams provide the major contribution.  
This is very remarkable, since the folklore picture of $\Phi$ photoproduction is that it rather happens via a vector-meson-dominance-like mechanism which is rather resembled by the Class~I diagrams~\cite{Laget}.
On the right panel of Fig.~\ref{fig:Phiproduction} our results are confronted with experimental data from 
SLAC~\cite{Ballam}, Daresbury~\cite{Barber} and JLab~\cite{Anciant} ($\bullet$, $\blacktriangle$, $\blacksquare$). 
Out of the three data sets only the SLAC data are directly taken at the energy of $p_\gamma^{(lab)} = 9.3\,\text{GeV}$. 
To increase our data base we have chosen the strategy to scale the Daresbury and JLab data, 
which were taken at lower energies but larger scattering angles, 
such that they smoothly extrapolate the SLAC data. 
One sees that our HSA results reproduce the angular dependence of the experimental cross-section data very well 
and have the right order of magnitude -- in contrast to other photoproduction channels. 
An increase of the magnitude of our predictions is even to be expected if one takes a more realistic, broader vector meson DA 
instead of the simple, non-relativistic delta function. 
It should also be mentioned that, since we have neglected light and strange-quark masses, 
the proton helicity has to be conserved and the $\Phi$ must be polarized longitudinally.
This means that helicity amplitudes where the $\Phi$ is transversely polarized do not contribute. 
\begin{figure}[b]
\subfigure{\includegraphics[width=0.45\textwidth, height=0.35\textwidth]{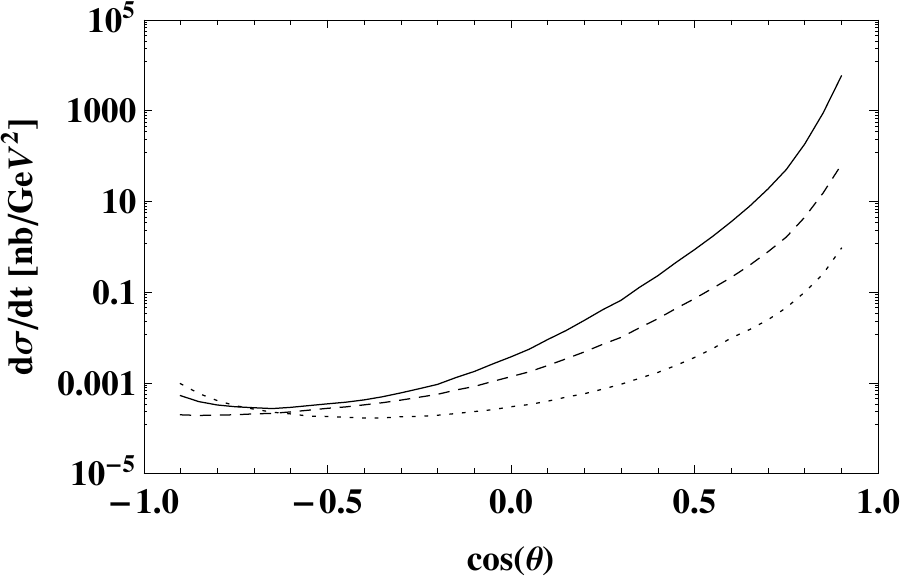}}
\hfill
\subfigure{\includegraphics[width=0.45\textwidth, height=0.35\textwidth]{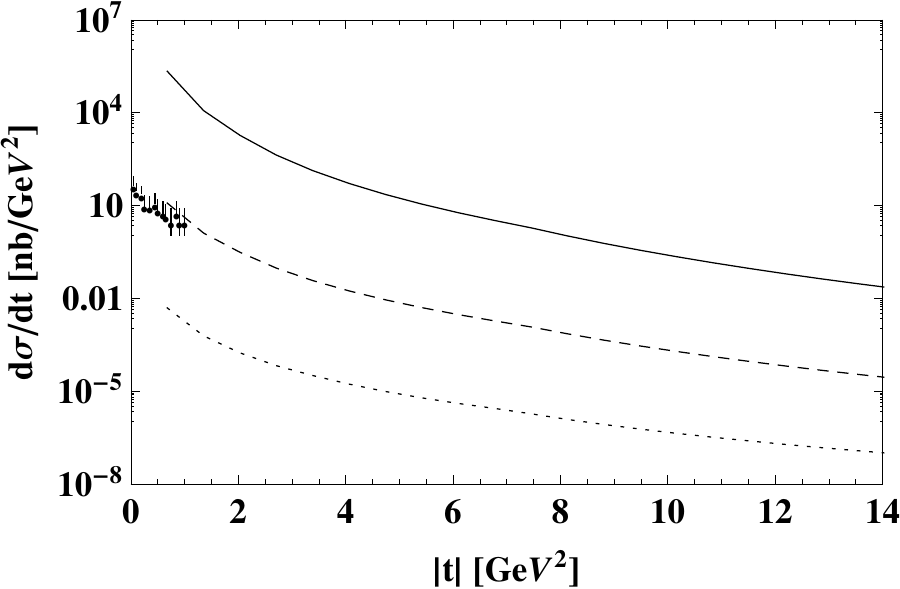}}
\caption{
 {\it Left:} The influence of the charm-quark mass on $d\sigma_{\gamma p \rightarrow J/\Psi p}/dt$ for
  $p_\gamma^{\left(lab\right)} = 20$~GeV.
  The cross section for $\lambda_{J/\Psi}=0$ without (dotted) and with (dashed-dotted) charm-quark mass
  is compared to the cross section for $\lambda_{J/\Psi} = \pm 1$ (solid line). 
 {\it Right:} Our predictions for $d\sigma_{\gamma p \rightarrow J/\Psi p}/dt$
  at $p_\gamma^{\left(lab\right)}=150$~GeV in comparison with FNAL data. 
  The cross section for $\lambda_{J/\Psi}=0$ without (dotted) and with (dashed-dotted) charm-quark mass
  is compared to the full cross section (solid line) that includes $\lambda_{J/\Psi}=\pm 1$ contributions.
} 
\label{fig:JPsiproduction}
\end{figure}

We now turn to the discussion of photoproduction of the heavier $c\bar{c}$ quarkonium $J/\Psi$.  
Including the heavy charm-quark mass in our calculation for $J/\Psi$ photoproduction leads to considerable differences 
as compared to $\Phi$ photoproduction, where the strange-quark mass has been neglected. 
On the left panel of Fig.~\ref{fig:JPsiproduction} the influence of the charm-quark mass on the 
unpolarized differential cross section $d\sigma_{\gamma p \rightarrow J/\Psi p}/dt$ 
at the photon lab energy $p_\gamma^{\left(lab\right)} = 20$GeV is shown.   
The dotted line represents our HSA prediction for the charm-quark mass set to zero. In this case only longitudinally polarized $J/\Psi$s
can be produced. 
The inclusion of the charm-quark mass leads already to a significant increase of the production cross section for longitudinally polarized $J/\Psi$s (dashed line). 
But, due to the finite charm-quark mass, the production of transversely polarized $J/\Psi$s becomes also allowed. 
As it turns out, transversely polarized $J/\Psi$s give even the dominant contribution to the $J/\Psi$-production cross section (solid line).
On the right panel of Fig.~\ref{fig:JPsiproduction} we compare experimental FNAL data \cite{Binkley} 
for $p_\gamma^{\left(lab\right)} = 150 \ \text{GeV}$ with our HSA predictions. 
Again the experiments have been performed only for low momentum transfer with $|t|$ up to $\approx 1.5$~GeV$^2$.  
As before, the dotted line shows our prediction for zero charm-quark mass, which by far underestimates the experimental data. 
Our result for the massive longitudinal case, represented by the dashed curve, seems to provide a very good extrapolation to the FNAL data. 
But taking also into account the helicity amplitudes where the $J/\Psi$ is transversely polarized worsens the situation. 
The full calculation of $d\sigma_{\gamma p \rightarrow J/\Psi p}/dt$, shown by the solid line, 
overshoots the FNAL data by nearly two orders of magnitude.  

The observation that \lq\lq mass effects\rq\rq\ dominate the $J/\Psi$ photoproduction cross section in the kinematic situations considered in Fig.~\ref{fig:JPsiproduction}  lets us conclude that both, $|t|$ and $m_{J/\Psi}$, still play  an important role as large scales and a description by means of the leading order HSA is insufficient. For the photoproduction of $\Phi$ mesons the situation looks much better. The angular dependence of the predicted cross sections agrees with the (sparse) experimental data and their absolute magnitude is of the right order. Remarkably, diagrams which resemble a Compton-scattering-like production mechanism dominate over diagrams constituting the remnants of a vector-meson-dominance-like mechanism. The 12 GeV upgrade of CEBAF at JLab will give us a good chance for an experimental answer to the question, whether a substantial amount of the $\gamma p \rightarrow \Phi p$ differential cross section at momentum transfers of a few GeV can come from a perturbative production mechanism as considered here.

\medskip
\noindent 
{\bf Acknowledgement:} 
This work was supported via an agreement for scientific and technological cooperation between Austria and Croatia 
(\"OAD project number 19/2004). 
A.T.G. acknowledges the support of the \lq\lq Fonds zur F\"orderung der wissenschaftlichen Forschung in \"Osterreich - FWF\rq\rq\ 
(project J 3163-N16).

\end{document}